# Novel Mining of Cancer via Mutation in Tumor Protein P53 using Quick Propagation Network

Ayad. Ghany Ismaeel, and Raghad. Zuhair Yousif

*Abstract*— There is multiple databases contain datasets of TP53 gene and its tumor protein P53 which believed to be involved in over 50% of human cancers cases, these databases are rich as datasets covered all mutations caused diseases (cancers), but they haven't efficient mining method can classify and diagnosis mutations patient's then predict the cancer of that patient. This paper proposed a novel mining of cancer via mutations because there is no mining method before offers friendly, effective and flexible predict or diagnosis of cancers via using whole common database of TP53 gene (tumor protein P53) as dataset and selecting a minimum number of fields in training and testing quick propagation algorithm which supporting this miming method. Simulating quick propagation network for the train dataset shows results the Correlation (0.9999), R-squared (0.9998) and mean of Absolute Relative Error (0.0029), while the training for the ALL datasets (train, test and validation dataset) have results the Correlation (0.9993), R-squared (0.9987) and mean of Absolute Relative Error (0.0057).

## I. INTRODUCTION

TUMOR protein P53 (produced by TP53 gene) is a sequence-specific transcription factor that acts as a large tumor suppressor in mammals. Disable the function of the tumor suppressor p53 is from one of the most common genetic alterations in human cancer, and close to half of all human tumors carry p53 gene mutations within their cells. Fig. 1 shows the worldwide distribution of cancers and P53 mutations [1].

Databases related to tumor protein P53 (TP53 gene) contain large amounts of data, and Regular techniques may not be helpful and impractical in such large volumes of data, so artificial intelligence techniques such as data mining to facilitate and improve the process of research and education, data mining is the process of analyzing the data by linking them with artificial intelligence techniques for inspection and search for specific information and useful in a large volume of data, and it is through the process of linking the analysis of this data and methods of artificial intelligence to become the most efficient in the inspection process, and this so-called Knowledge Discovery in Database KDD.

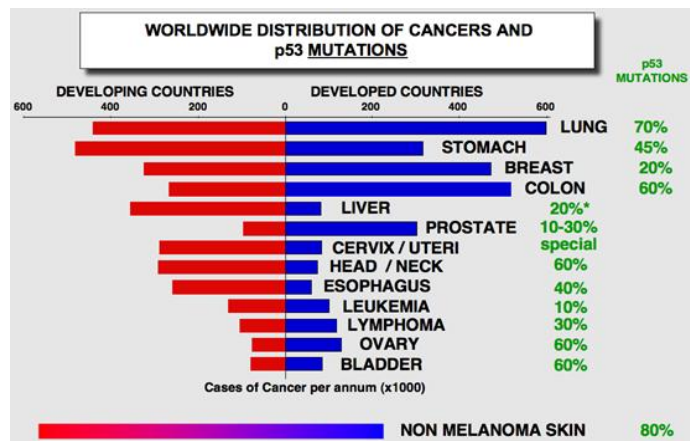

Fig. 1 Shows P53 (TP53 gene) mutations

Artificial Neural Networks ANNs learn from the examples that must be carefully selected in order to function correctly, ANN can discover how to solve problems on their own. There is needed for mining method focuses on mutations of tumor protein P53 these mutations used in train BPN, and then this trained BPN used to classify and diagnosis cancer via mutations in patient's P53 (TP53) sequence. This method supposed previously classifying of malignant mutations in patient's P53 using BioEdit packages summarized as follows [2, 3, and 4]:

1) Using person's TP53 sequence which obtained from oncogene labs in FASTA format to reach to normal homology of TP53 gene via NCBI using BioEdit as show in Fig 2.
2) Open BioEdit package with the normal gene of TP53 (Fasta file) to determine GC% whether equal or greater than 38%, otherwise search another normal gene of TP53, while the obtained normal gene of TP53 shows CG%= 54.85% continue to next.
3) Using that fasta file of normal gene of TP53 in ClustalW of BioEdit package to classify alignment result, i.e. diagnosis there is malignant mutations by comparing the normal TP53 gene sequence with one (or more than one) person's TP53 gene sequences at the same time, e.g. a mutation in person's TP53 gene comparing with normal TP53 gene as shown in Fig. 3.

Ayad. Ghany Ismaeel, is Professor in dept. of Information Systems Engineering, Technical Engineering College, Erbil Polytechnic University, IRAQ. (phone: 009647504292103; e-mail: dr.ayad.ghany.ismaeel@gmail.com).

Raghad. Zuhair Yousif, is currently Professor, Assistant at branch of Communication in Department of Applied Physics College of science at Salahaddin university-Erbil, IRAQ. (e-mail:drraghad.zuhair@gmail.com ).

International Journal of Computer Science and Electronics Engineering (IJCSEE) Volume 3, Issue 2 (2015) ISSN 2320–4028 (Online)

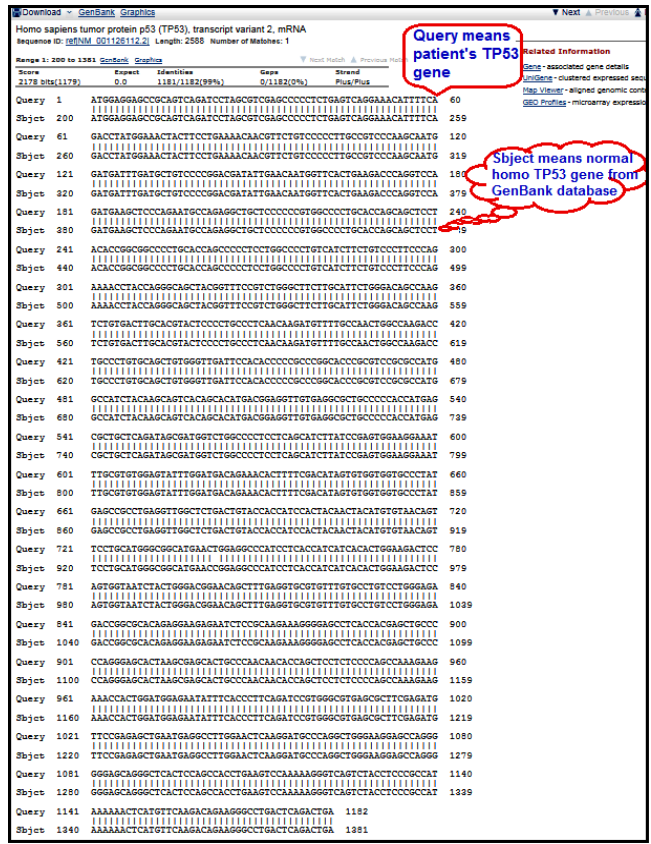
Fig. 2 The normal homology TP53 gene

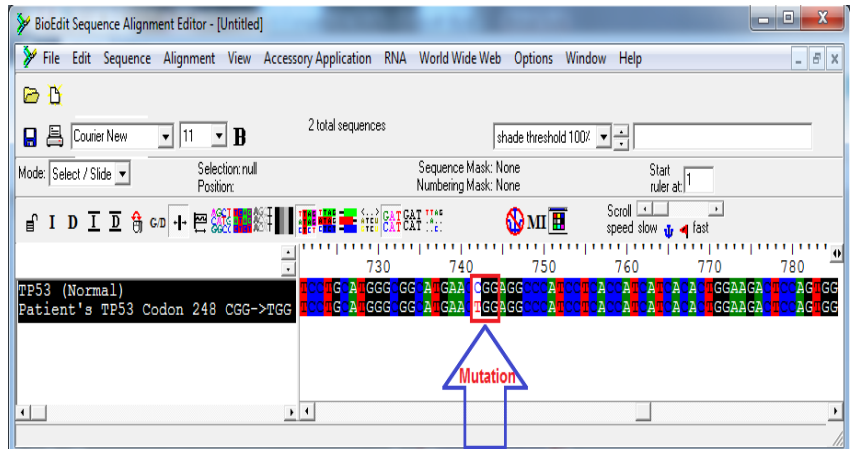
Fig. 3 Shows mutation at person's TP53 comparing with normal TP53 gene

Then transform normal homology TP53 gene and person's TP53 gene to tumor protein P53 (depending on the idea of "two sequences may have big differences in DNA sequence but have similar protein") using same ClustalW of BioEdit package to diagnosis there is malignant mutations or not (no risk). Fig. 4 shows there is malignant mutations at codon 248 (CGG→ TGG), i.e. will find the codon 248 converted from R (in normal P53) to W (in Person's P53 gene).

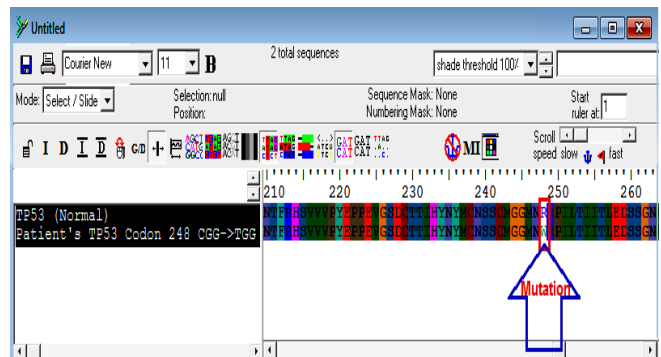
Fig. 4 Shows there are malignant mutations at codon 248 (CGG→TGG)

Classified malignant mutations which are discovered at step (4) needed to diagnose and predict a cancer or pre-cancer for Person using mining method by determines the cancer(s) related to the classified mutation, as know there are multiple cancers related with each mutated codon in tumor protein P53. The mining method which is needed base on training Quick Propagation Network QPN which is an improvement of the back propagation network (QPN means converges faster than back propagation network). QPN based on Newton's method to estimate the solution, the method itself requires knowing a second derivative of the error function [7].

## II. RELATED WORK

E. Adetiba, J. C. Ekeh, V. O. Matthews, and et al. [2011], proposed study aims estimating an optimum back-propagation training algorithm for a genomic based ANN system for NSCLC diagnosis, the nucleotide sequences of EGFR's exon 19 of a noncancerous cell were used to train an artificial neural network (ANN). Several ANN back propagation training algorithms were tested in MATLAB R2008a to obtain an optimal algorithm for training the network, in the nine different algorithms tested and achieved the best performance (i.e. the least mean square error) with the minimum epoch (training iterations) and training time using the Levenberg-Marquardt algorithm [5].

S. Sathish Kumar and Dr N Duraipandian [2013], proposed Artificial Neural Network technique to classify the disease with reduced number of DNA sequence. The accuracy is differing based on the training data set and validation data set. The other major issue is the privacy preserving of the patients. This proposed will share the critical data from clinical diagnostic centers, there is good chance of patient's anonymity is revealed. To avoid this used a simple Privacy Preserving in Data Mining (PPDM) technique to crypt the identity of the patients as well as the critical data and discloses only the required data like DNA sequence to the research team. The proposed technique effectively identifies the gene from its gene sequence and so the disease moreover, results obtained from 10-fold cross validation have proved that the disease can be identified even from a part of the DNA sequence. Though the technique has been tested with the DNA sequence of only two genes (PIK3CA and TP53), the 10-fold cross validation results have reached a remarkable performance level. From the results, it can be hypothetically analyzed that a technique, which identifies the disease only with a part of gene sequence, has the ability to classify any kind of disease [6].

The drawbacks of these methods and techniques focus in diagnostic or prediction base on reduce number of sequence as clinical data (i.e. not base on whole database of genes caused diseases) and the dataset not big enough like whole database to give results enough. The motivation overcomes the drawbacks of the previous techniques to reach a novel mining method of cancers via mutations in beg dataset (whole database) of tumor protein P53 using quick propagation network.

## III. PROPOSED A NOVEL MINING METHOD OF CANCERS VIA MUTATIONS IN TUMOR PROTEIN P53

The proposed mining method contains two stages as follow:
1) Learning or training Quick Propagation Network:
   The proposed structure of training QPN as an improvement of the back propagation network [3, 7] has 2 layers (input layer not within layers of QPN). The formula of the Newton's method which is supported QPN as follow:

$$w_{n+1} = w_n - \frac{J'(w_n)}{J''(w_n)} \quad (1)$$

The geometric interpretation of the proposed mining method an approximate at each iteration the function (*J*) by a quadratic function around (*W*), and then makes a step towards the minimum of that quadratic function. In order to avoid calculating a second derivative by ourselves we could use a secant method to numerically compute the second derivative approximation, here is how it looks:

$$J''(w_n) \approx \frac{J'(w_n) - J'(w_{n-1})}{w_n - w_{n-1}} \quad (2)$$

Combining the two equations can get such formula:

$$w_{n+1} = w_n - \frac{J'(w_n)}{\left(\frac{J'(w_n) - J'(w_{n-1})}{w_n - w_{n-1}}\right)}$$
$$= w_n - \frac{J'(w_n)}{J'(w_n) - J'(w_{n-1})}(w_n - w_{n-1})$$
$$= w_n + \frac{J'(w_n)}{J'(w_{n-1}) - J'(w_n)}(w_n - w_{n-1})$$
$$(3)$$

Replacing the weights increment by a delta symbol will get formula, known as QPN:

$$\Delta w_n = w_n - w_{n-1}$$
$$\Delta w_{n+1} = w_{n+1} - w_n$$
$$\Delta w_{n+1} = \frac{J'(w_n)}{J'(w_{n-1}) - J'(w_n)} \Delta w_n$$
$$(4)$$

2) Test of Quick Propagation Network to Diagnose Cancers:
   The testing (via query) using training QPN in (1) above will diagnose the type of cancer for classified mutation of patient's P53 sequence at section 1; step 4. The algorithm of testing QBP to diagnose and predict cancers via mutations in Person's P53 sequence as follow:

**Input:** Patient's P53 gene sequence.**(As referring to in section1;steps1-4)**

**Output**: Diagnose the type of classified mutation in Patient's P53 sequence.

**BEGIN**
   If there is mutation
     If manually:
       Input part of fields required in testing (Query)
     Else
       Using file to input part of fields required in testing (Query) for more than Patients
     The novel mining will retrieve the output fields include cancer.
   Else
     No risk (no cancer retrieved)
**END**

## IV. SIMULATION RESULTS

The proposed novel mining of cancers via mutations in tumor protein P53 simulated using ANI ver. 2.1 as follow.

### A. Learning or Training Quick Propagation Algorithm

The dataset used to train QPN, this dataset (all records = 1447) of common database (UMD_Cell_line_2010) from TP53 website as modern and comprehensive database under URL:

http://p53.free.fr/Database/p53_MUT_MAT.html[1].

Effective fields (part of record) selected from dataset and the target field which selected (Mutation position) using Alyuda NeuroIntelligence ANI as simulation package. ANI can select an ideal train set TRN, validation set VLD and test set TST to train and test QPN, sample of this dataset shown in Fig 5.

Fig. 5 shows sample of dataset (database) which used in training QPN

To optimize the modeling, the results of algorithms related to miming method must be reached to:
1) R-squared: Statistical ratio that compares model forecasting accuracy with accuracy of the simplest model that just use mean of all target values as the forecast for all records. When R-squared become closer to (1) the better the model is.
2) Correlation: Statistical measure of strength of the relationship between the actual values and network outputs. The r coefficient can range from -1 to +1; closer r is to 1, the stronger positive linear relationship.
3) Absolute Relative Error ARE: is an error value that indicates the "quality" of the neural network training. This index is calculated by dividing the difference between actual and desired output values by the module of the desired output value.

Optimal design of QPN with its topology means number of nodes needed in input, hidden and output layers. Alyuda NeuroIntelligence gives (283-141-1) as shown in Fig. 6.

Fig. 7 shows training QPN with topology (283-141-1) Absolute error in training (1.9744) while in validation (6.9706), Network error in training (0.000006) while in validation (0) and Error improvement is (9.27E-09). The result of training QPN (Actual vs. Output) in case of TRN set shown in Fig. 8.

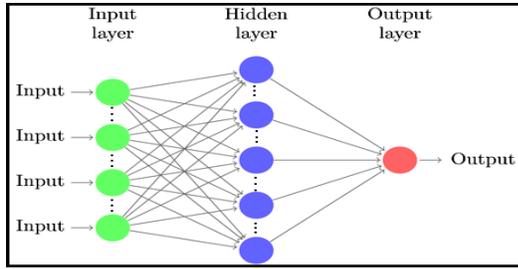

Fig. 6 Topology (283-141-1) of QPN with 2 layers Structure

TABLE I
SHOWS THE RESULTS OF TRAINING QPN IN DIFFERENT TYPES OF DATASETS

| Dataset Type | Correlation | R-squared | Mean of ARE |
|---|---|---|---|
| Train set TRN | 0.9999 | 0.9998 | 0.0029 |
| Validation set VLD | 0.9979 | 0.9957 | 0.0126 |
| Test set TST | 0.9979 | 09956 | 0.0128 |
| ALL sets | 0.9993 | 0.9987 | 0.0057 |

### B. Testing QPN Algorithm to Diagnose Cancers

Based on the algorithm referring in section 3, step 2 predicts and diagnosis cancers via mutations of Person's P53 sequence using testing QPN which trained at subsection 4.1. The proposed mining method allows predict cancers via mutations of certain person (by entering each field parameter manually) or group of persons (loading the parameters for group of persons once via file), as shown in Fig. 9.

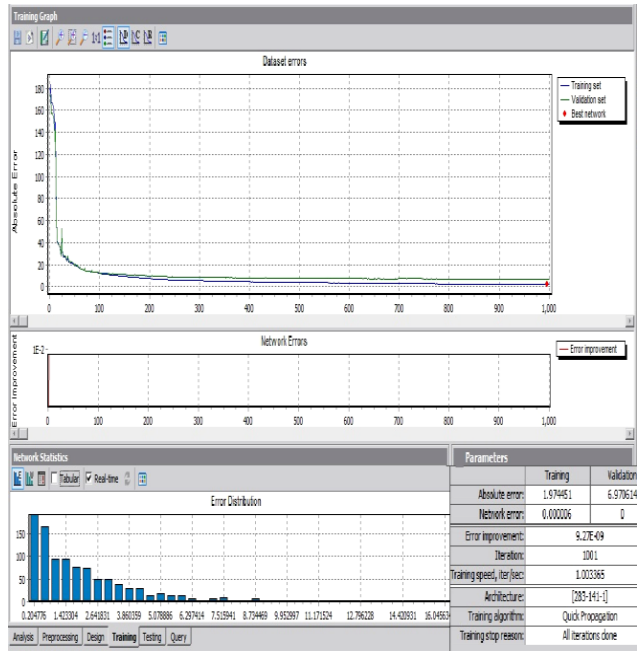

Fig. 7 Shows result of training QPN with topology (283-141-1)

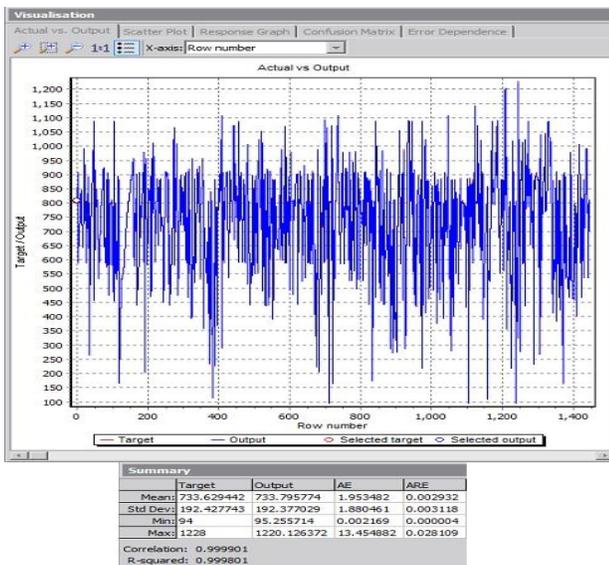

Fig. 8 Shows result of training QPN in case of TRN dataset

Table 1, shows the results (Correlation, R-squared and Mean of ARE) of training QPN in different types of datasets.

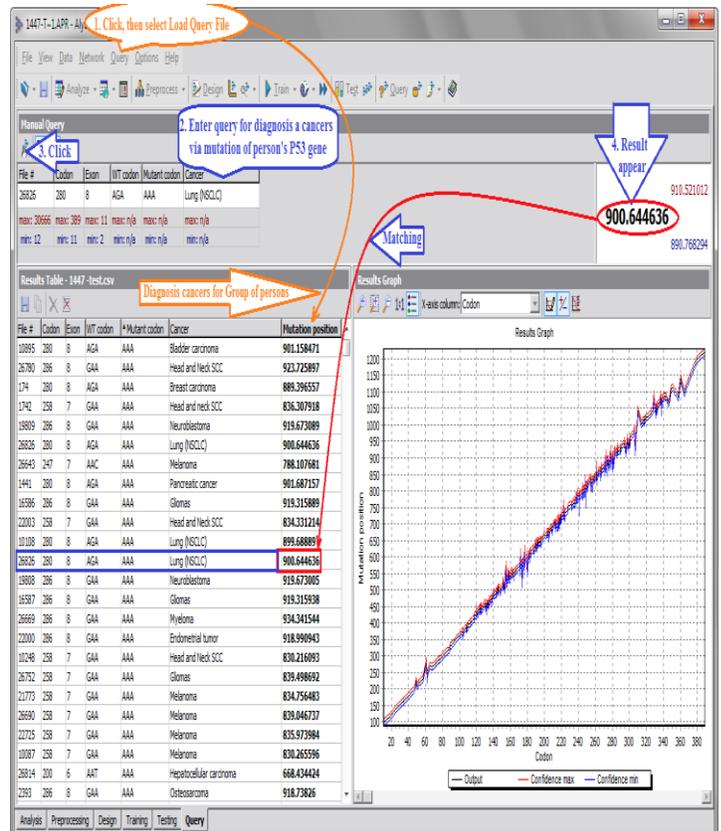

Fig. 9 Shows predict cancers via mutations of Person's P53 or group of Persons

### C. Discussion the Results

Table 2 show comparing the results of proposed mining method of predict cancer via detecting mutated P53 with other related techniques or methods.

TABLE II
REVEALS COMPARISON OF PROPOSED MINING METHOD WITH OTHER TECHNIQUES OR METHODS

| Proposed Mining Method | E. Adetiba, J. C. Ekeh, V. O. Matthews, and et al.[5] | S.Sathish Kumar and Dr N Duraipandian[6] |
|---|---|---|
| Using quick propagation network, i.e. improvement (faster than) BPN | Standard Backpropagation network | Standard Backpropagation network |
| Beg dataset (whole database), while the mining method using part of each record (effective fields). | Small dataset comparing to proposed mining method | Small dataset comparing to proposed mining method |
| Training Dataset based on TP53 gene caused more than 50% of human cancers | Another gene sequence | TP53 and PIK3CA gene sequence |
| This proposed mining method based on classifying malignant mutations then diagnoses the cancers related. | There isn't this feature | There isn't this feature |
| The results of proposed mining method more effective | Not like the proposed mining method | Not like the proposed mining method |

## V. CONCLUSIONS

The proposed mining method of cancer via mutations in tumor protein P53 shows the following conclusions:

1) This proposed method is a novel mining cancer via mutation in tumor protein P53, because there is no mining method before offers friendly diagnosis and detecting malignant mutations caused cancers. Addition it works on whole database of TP53 gene as referring to in subsection 4; A.

2) The mining method is effective in diagnosis and predicts cancers via mutation, where it used in training and testing (query) QPN a minimum number of fields (6) fields from (53) fields in each record of TP53 database [at URL: *http://p53.free.fr/Database/p53_MUT_MAT.html*], as shown in Fig. 5.

3) Mining method which is proposed allows flexible diagnosis and predicts cancers via mutations in tumor protein P53 sequence via two options first one manually query to diagnosis cancer for one person while the second for group of persons once by loading file contains the fields (parameters) which are used in diagnosis cancers of this group of persons.

## ACKNOWLEDGMENT

Thank to (DR. NEMA SILAH ABDAL KAREEM) that helped in the maturation of the idea of research and advices at the medical side, which has to do with their competence.

**Ayad Ghany Ismaeel**: Received Computer Science PhD (Computer and IP Network) from Technology University, Baghdad, IRAQ at 2006 and M.Sc. in computer science (Applied) from NCC, ministry of Higher Education and Scientific Research, IRAQ at 1987. At 1981/1982 he obtained B.Sc. (Bachelor of Science) in Statistics/Informatics - University of Al'mustansaryha, Baghdad– Iraq.
He is participated multiple workshops and courses sponsored by GIZ-Germany, UK, JICA-Japan, etc. which offered to him experience and support in Advising, Counseling, Teaching, Training and Curriculum Development. Dr. Ayad is professor in postgraduate (doctorate, master, and higher diploma) and undergraduate studies in computer science, information systems engineering and software engineering at the universities of Baghdad and Kurdistan region, IRAQ.
Professor Ismaeel is currently in dept. of Information Systems Engineering, Technical Engineering College, Erbil Polytechnic University, IRAQ. He is editorial board, program committee member and board reviewer for different international journals and conferences. His research interests in mobile, IP networks, web application, GPS & GIS techniques, distributed databases, information systems, bioinformatics and bio-computing. He has many publications in international and national journals and conferences, supervising M.Sc. and higher diploma degree projects in software engineering and computer science, also participated in different discussions, examination committees of PhD and M.Sc. More details visit his website: http://drayadghanyismaeel.wix.com/ayad-ghany-ismaeel-

**Raghad Zuhair Yousif:** received B.Sc. in Electronics and Communication Engineering from Baghdad University College of Engineering Department of Electronics and Communication Engineering in 1998. Then he received MSc. In Electronics and Communication Engineering from Al-Mustansriyha University in Baghdad College of Engineering Department of Electrical Engineering in 2001. His research was in field of image processing and data security. Then he received a PhD. in Communication Engineering form Department of Electrical and Electronic Engineering from Baghdad University of Technology in 2006.
His research was in field of FPGA and channel coding. He had been worked as senior lecturer at Department of Software Engineering College of Engineering Salahaddin University–Hawler from 2006 to 2010. He is currently Professor Assistant at branch of Communication in Department of Applied Physics College of science at Salahaddin university- Hawler. His research in areas of interest are Reconfigurable hardware, Channel Coding, Real Time Systems, Medical Image processing, Remote sensing, Data Security, Bioinformatics and Computer Networks. He senior lecturer at many MSc Courses for remote sensing, Advanced Computer Networks, Multimedia Technology, Network Security, Real time systems, and supervisor of Many MSc. Thesis in Software Engineering.